\documentclass[aps,twocolumn,a4paper,showpacs]{revtex4}
\usepackage{graphicx}
\usepackage{amsmath,amssymb}
\usepackage{enumerate}
\usepackage{subfigure}
\usepackage{tabularx}

\newcommand{\be}{\begin{equation}}
\newcommand{\ee}{\end{equation}}
\newcommand{\ben}{\begin{eqnarray}}
\newcommand{\een}{\end{eqnarray}}
\newcommand{\bes}{\begin{subequations}}
\newcommand{\ees}{\end{subequations}}

\newcommand{\bb}{\bibitem}

\begin{document}
\title{Semi-Compact Skyrmion-like Structures}
\author{D. Bazeia and E.I.B. Rodrigues}
\affiliation{Departamento de F\'\i sica, Universidade Federal da Para\'\i ba, 58051-970 Jo\~ao Pessoa, PB, Brazil}
%%%%%%%%%%%%%%%%%%%%%%%
\begin{abstract}
We study three distinct types of planar, spherically symmetric and localized structures, one of them having non-topological behavior and the two others being of topological nature. The non-topological structures have energy density localized in a compact region in the plane, but are unstable against spherically symmetric fluctuations. The topological structures are stable and behave as vortices and skyrmions at larger distances, but they engender interesting compact behavior as one approaches their inner cores. They are semi-compact skyrmion-like spin textures generated from models that allow to control the internal behavior of such topological structures.
\end{abstract}
\date{\today}
\maketitle
%%%%%%%%%%%%%%%%%%%%%%%%%%%%%%%%%%
\section{Introduction}

The study of skyrmions first appeared in the context of particle physics \cite{sky} and have been extended to various other areas of nonlinear science, in particular as localized structures in a diversity of magnetic materials; see e.g., Refs.~\cite{yu,seki,fert,a1,Bog1,new1,new2,Bog2,oleg}. In this work we focus mainly on magnetic skyrmions, dealing with some analytical tools to describe the formation of skyrmions in magnetic materials. The subject concerns several mechanisms working together \cite{nag} and describes  properties such as topological charge or winding number, vorticity and helicity \cite{zhang}. In general, the topological charge of a skyrmion may be integer, $Q=\pm 1$, half-integer, $Q=\pm 1/2$ (half-skymion or vortex) \cite{oleg,young,MV0,MV1,MV2,MV3}, and null $Q=0$, the later being non-topological, considered in Ref.~\cite{a3}. Here one notes that skyrmions with high-topological-number ($Q\geq2$) has also been suggested; see, e.g., the recent work \cite{ezawa16}.

We will deal with skyrmion-like structures, paying closer attention to compact solutions,
as one explains below. Compact skyrmions have been studied before in \cite{S}, in the context of the baby Skyrme model with a specific potential, and also in \cite{E}, to describe magnetic spin textures that appear in magnetic materials. The study of compact skyrmions in magnetic materials is of current interest, since the recent advances on miniaturization is leading us to manipulate magnetic materials at constrained geometries and this may modify the profile and properties of the magnetic spin textures that inhabits the material. An example of this was identified experimentally in \cite{DW}, showing how a domain wall may change profile as it is shrank inside a constrained region. For this reason, we think it is worth investigating the possibility of shrinking vortices and skyrmions to a compact region in the plane.

In the current work, we follow the lines of Refs.~\cite{7,jmmm,8} and use scalar fields to model spin textures for which we can control their spin arrangements. The investigation presented in \cite{E} has found periodic solutions described in terms of the Jacobian elliptic functions, which are then worked out to give rise to compact structures. Here we follow another route, inspired by the recent investigations \cite{lump,as} on compact and asymmetric structures of current interest to high energy physics. We then introduce models that provide semi-compact solutions, which are new isolated structures that engender integer and half-integer skyrmion numbers, for which we are able to control the spin textures at their inner core.

The study deals with three distinct models, one describing non-topological solutions, with topological charge $Q=0$, and the two others being able to generate topological structures with integer and half-integer topological charge. In the case of solutions with vanishing topological charge, we show that they are linearly unstable against radial deformations, so they are of limited interest. But we deal with them because one can develop mechanisms to make them stable, although this is out of the scope of the current work. The solutions with integer and half-integer topological charge are linearly stable against radial deformations, so we focus mainly on them, showing how to control their inner core, without modifying their asymptotic behavior. We call them semi-compact structures, and they have an important difference, when compared to the compact structures discussed in Ref.~\cite{E}. The key point here is that the semi-compact solutions that we introduce have asymptotic behavior similar to the standard spin textures, but they behave differently as one approaches the core of the structure. We understand that this is important behavior, because it help us show how to control the inner core of skyrmions.

We organize the work as follows: in the next section we follow Refs.~\cite{7,jmmm} and discuss three distinct models describing a single real scalar field in two spatial dimensions, as suggested in Ref. \cite{8}. We describe explicit solutions, and study their stability. In Section III, we deal with the topological charge density and study the topology and the skyrmion number. We end the work in Sec. IV, where we include our comments and conclusions.

%%%%%%%%%%%%%%%%%%%%%%%%%%
\section{Analytical procedure} 

We start focusing on the localized spin textures that appear in magnetic materials. We suppose that the material is homogeneous along the ${\hat z}$ direction, with the magnetization ${\bf M}$ being in general a three-component vector with unit modulus that depends on the planar coordinates, such that ${\bf M}={\bf M} (x,y)$ and ${\bf M}\cdot{\bf M}=1$.

To describe skyrmions, we introduce the skyrmion number, which is a conserved topological quantity, given by
\be\label{Q}
Q=\frac{1}{4\pi}\int_{-\infty}^{\infty}\!\!\! dx\, dy \;{\bf M}\cdot\partial_x{\bf M}\times\partial_y{\bf M}.
\ee
In this work we concentrate on helical excitations, with the magnetization ${\bf M}={\bf M}(r)$ only depending on the radial coordinate, being now a two-dimensional vector, orthogonal to the radial direction. In cylindrical coordinates, the helical excitations impose that ${\bf M}\cdot{\hat r}=0$, so one allows that the magnetization has the form
\be\label{M}
{\bf M}(r)={\hat \theta}\cos\Theta(r) + {\hat z}\sin\Theta(r),
\ee 
With this decomposition, we can then describe $\Theta(r)$ as the only degree of freedom to be used to model the magnetic excitation in the magnetic material. If we now use Eq.~\eqref{Q} with the $(r,\theta)$ variables, the above magnetization can be used to obtain the skyrmion number. It leads to 
\be\label{q}
Q=\frac12\sin\Theta(0)-\frac12\sin\Theta(\infty).
\ee
This result shows that the topological profile of the solution $\Theta(r)$ is related to its value at the origin, and the asymptotic behavior for $r\to\infty$.

In order to model skyrmion-like solutions analytically, we take advantage of the recent study \cite{7,jmmm} and consider
\be\label{T}
\Theta(r)=\frac{\pi}2\phi(r)+\delta,
\ee
where $\delta$ is a constant phase, which we use to control the value of the magnetization at the center of the magnetic structure. Also, we suppose that the scalar field $\phi$ is homogeneous and dimensionless quantity which is described by the planar system investigated in \cite{8}. In this case, the Lagrange density ${\cal L}$ has the form
\be\label{model}
{\cal L}=\frac12\dot\phi^2-\frac12 \nabla\phi\cdot\nabla\phi-U(\phi),
\ee
where dot stands for time derivative, and $\nabla$ represents the gradient in the $(x,y)$ or $(r,\theta)$ plane. We search for time independent and spherically symmetric configuration, $\phi = \phi(r)$, and consider $U = U(r;\phi)$ in the form
\be\label{modelU}
U(r,\phi)=\frac1{2r^2}P(\phi),
\ee
with $P(\phi)$ an even polynomial which contains non-gradient terms in $\phi$. In this case, the field equation becomes
\be\label{modeleq}
r^2\frac{d^2\phi}{dr^2}+ r \frac{d\phi}{dr}-\frac{1}{2}\frac{dP}{d\phi} = 0,
\ee
and the energy for static solution $\phi(r)$ is given by
\be
E=2\pi\int_0^\infty rdr\rho(r),
\ee
with $\rho(r)$ being the energy density, such that
\be
\rho(r) = \frac12 \left(\frac{d\phi}{dr}\right)^2+ \frac{1}{2r^2}P(\phi).
\ee 

The point here is that we can describe explicit models, using distinct functions for the Polynomial $P(\phi)$. We do this for three distinct models in the next subsections. Before doing that, however, we note that for $\phi\to\phi_v$, with $\phi_v$ being constant and uniform, the energy density vanishes if $\phi_v$ is a minimum of $P(\phi)$ such that $P(\phi_v)=0$ \cite{7}. In this case, $\phi_v$ is a ground state and will guide us toward the construction of models.

Another issue is that the model one uses engender scale invariance; see, for instance, the equation of motion \eqref{modeleq}. An interesting route to describe magnetic skyrmions is to break scale invariance, and this can be achieved via the introduction of Dzyaloshinskii-Moriya interactions (see, e.g., \cite{ezawa16} and references therein); another route is to describe skyrmions on a lattice (see, e.g., \cite{lat} and references therein). In this work we will keep using the model \eqref{model} to describe the magnetization since it will lead to analytical solutions. In this case, although we cannot use results to describe quantitative properties, we can still use them qualitatively and, more importantly, we can explore the topological properties of the solutions, because the topology follows from their asymptotic behavior, and may appear despite the breaking of the scale invariance.

%%%%%%%%%%%%%%%%%%%%%%%
\subsection{Non-topological solutions} 

\begin{figure}[t!]
\centerline{\includegraphics[scale=0.4]{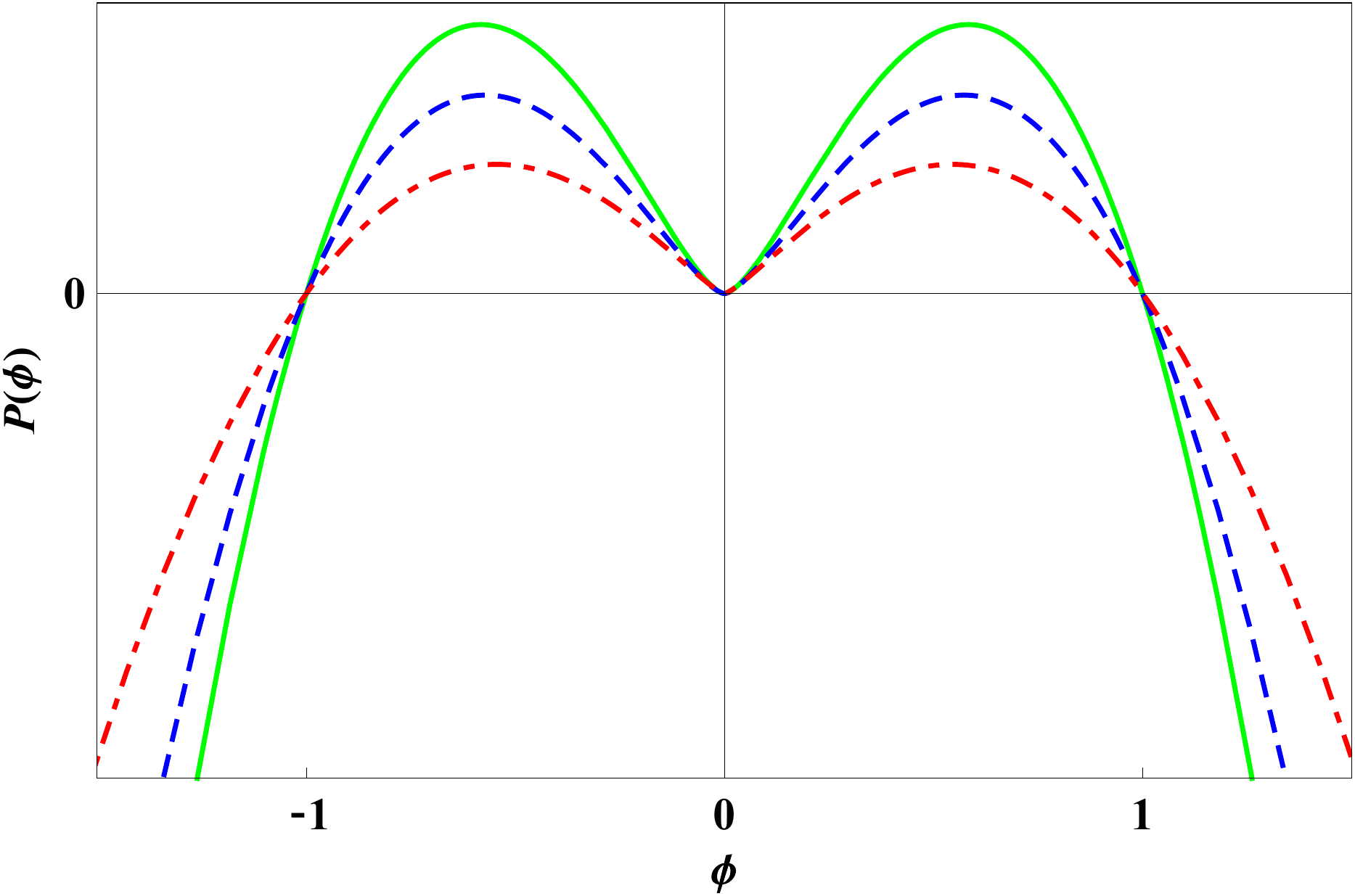}}
\caption{The polynomial \eqref{p2}, depicted for $n=3, 5,$ and $7$, with dot-dashed (red), dashed (blue) and solid (green) lines, respectively.}
\label{fig1}
\end{figure}

\begin{figure}[t!]
\centerline{\includegraphics[scale=0.4]{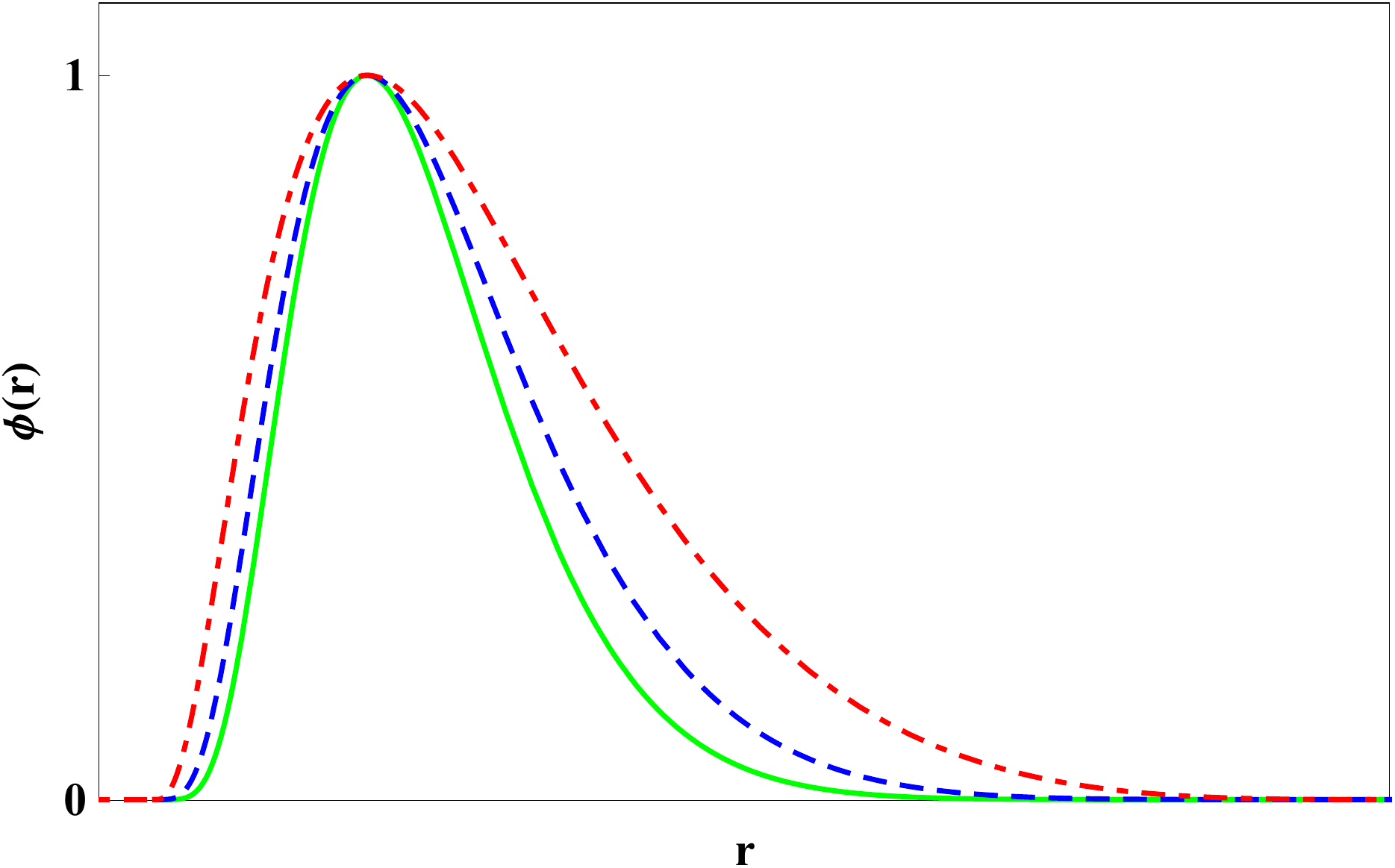}}
\caption{The solution \eqref{phi2} for the model \eqref{p2}, depicted for $n=3, 5$ and $7$, with dot-dashed (red), dashed (blue) and solid (green) lines, respectively.}
\label{fig2}
\end{figure}

The first model that we describe is governed by the polynomial $P(\phi)$, given by
\be\label{p2}
P(\phi)=n^2\phi^2(\phi^{-2/n}-1).
\ee
Here $n=3,5,7,...$ is an odd integer. This model follows the suggestion of Ref.~\cite{lump}, concerning the presence of one-dimensional, compact lump-like structures there studied. We depict this polynomial $P(\phi)$ in Fig.~{\ref{fig1}} for three distinct values of $n$, to show how it behaves as one changes such parameter. We see that the polynomial has a local minimum at $\phi_0 = 0$, and two zeroes at $\phi_\pm = \pm 1$, irrespective of the value of $n$. We can write the equation of motion as
\be
r^2\frac{d^2\phi}{dr^2}+ r \frac{d\phi}{dr}+n^2\phi-n(n-1)\phi^{1 -2/n }=0.
\ee
One then searches for solutions related to the ground state $\phi_0=0$. The investigation has led us with the interesting family of analytical solutions 
\begin{eqnarray}
\label{phi2}
\phi_n(r)=\left\lbrace
\begin{array}{ll}
\ \displaystyle \text{cos}^n(\text{ln}(r)), \quad e^{-\pi/2} \leq r \leq e^{\pi/2} \\ 
\ \\
\  0, \quad r < e^{-\pi/2} \quad \text{or} \quad r > e^{\pi/2} \\ 
\end{array}
\right..
\end{eqnarray}
There is another solution, with the minus sign, but it behaves similarly. The profile of the above solutions are depicted in Fig. \ref{fig2}, for three values of $n$. We note that as $n$ increases, the solution becomes more and more localized inside a compact region. We also note that the solution goes to zero at the origin and asymptotically, for $r\to\infty$, and this induces its
non-topological profile. This is an example of a compact solution that will be mapped into a compact skyrmion with vanishing topological charge, with vanishing skyrmion number. 

We use the above solution to see that the energy density is given by
\be
\rho(r)= \frac{ n^2}{r^2}\text{cos}^{2n}(\text{ln}(r))\left(\text{cos}^{-2}(\text{ln}(r))-1\right),
\ee
in the interval $e^{-\pi/2} \leq r \leq e^{\pi/2}$; also, it vanishes outside this compact region. 

We can integrate the energy density to get the total energy in the form
\be
E_n= \frac{n\pi^{3/2} \Gamma(n-1/2)}{\Gamma(n)},
\ee
where $\Gamma (z)$ is the Gamma function. In this case, one can use Eqs.~\eqref{q} and \eqref{T} to calculate the skyrmion number, giving $Q=0$. 

%%%%%%%%%%%%%%%%%%%%%%%%%%
\subsection{Topological solutions}

\begin{figure}[t!]
\centerline{\includegraphics[scale=0.4]{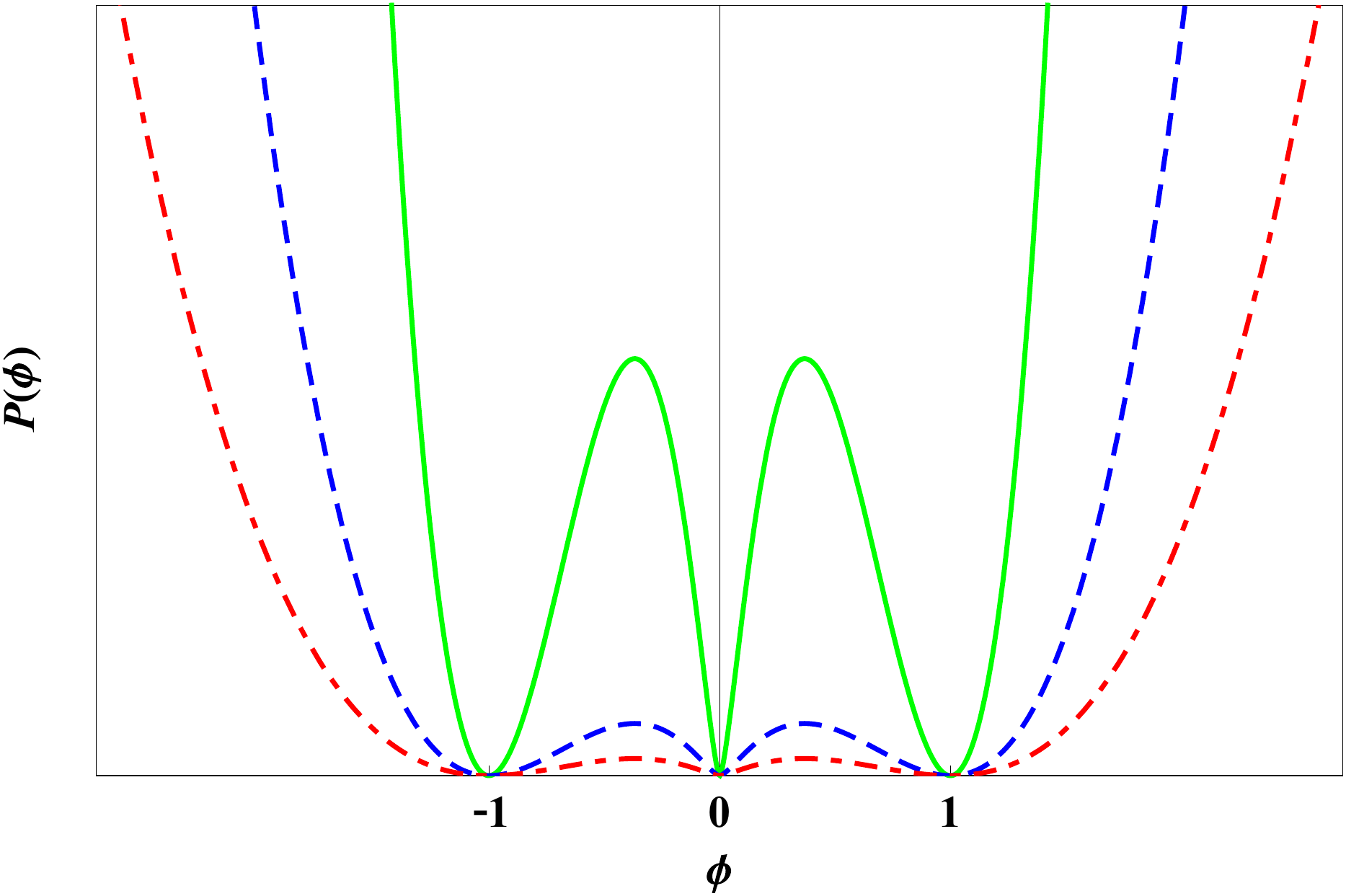}}
\caption{The polynomial \eqref{p1}, depicted for $s=0.3, 0.6$ and $0.9$, with dot-dashed (red), dashed (blue) and solid (green) lines, respectively.}\label{fig3}
\end{figure}

The second model is described by the polynomial $P(\phi)$ that has the form

\be\label{p1}
P(\phi)=\frac{1}{(1-s)^2}\phi^2 \text{ln}^2(\phi^2),
\ee
where the parameter $s$ is now in the interval $s \in [0, 1)$. This model has three minima, two at
$\phi_\pm = \pm 1$ and one at $\phi_0 = 0$. It was introduced in \cite{as}, and here we depict the above polynomial in Fig.~\ref{fig3}, for three values of $s$, to see how it changes as one varies $s$. 

%%%%%%%%%%%%%%%%%%%%%%%%%%%%%%%%%%%%%%%
%%%%%%%%%%%%%%%%%%%%%%%%%%%%%%%%%%%%%%%%
%%%%%%%%%%%%%%%%%%%%%%%%%%%%%%%%%%%%%%%
%%%%%%%%%%%%%%%%%%%%%%%%%%%%%%%%%%%%%%%%

\begin{figure}[t!]
\centerline{\includegraphics[scale=0.4]{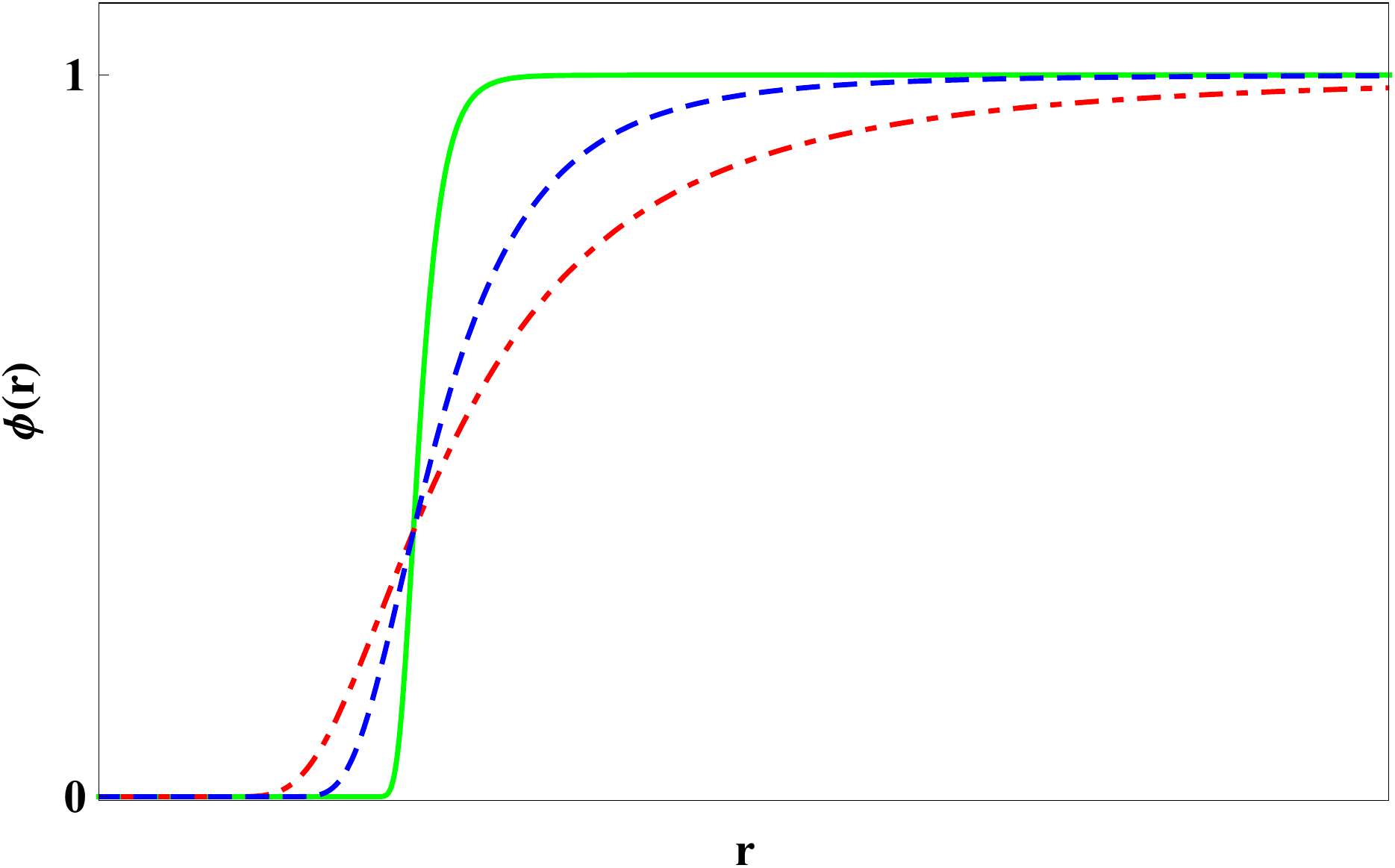}}
\caption{The solution \eqref{phi1} for the model \eqref{p1}, depicted for $s=0.3, 0.6$ and $0.9$, with dot-dashed (red), dashed (blue) and solid (green) lines, respectively.}
\label{fig4}
\end{figure}

We can write the equation of motion for static solutions as
\be
r^2\frac{d^2\phi}{dr^2}+ r \frac{d\phi}{dr}-\frac{\phi \text{ln}(\phi^2)(2+\text{ln}(\phi^2))}{(1-s)^2}=0,
\ee
and now one searches for solutions connecting the ground states $\phi_0=0$ and $\phi_+=1$; the equation can be solved analytically to give
\be\label{phi1}
\phi_{s}(r)= e^{-r^{-2/(1-s)}},
\ee
There is another solution, with the minus sign, which behaves similarly. We depict the solution (\ref{phi1}) in Fig. \ref{fig4}, for some values of s. We note that the solution is sharper for larger values of $s$. Also, it behaves standardly for very larger values of $r$, but it decays very strongly as $r$ approaches the center of the solution, indicating its semi-compact behavior. It vanishes for $r$ going to zero, and approaches the minimum $\phi_+=1$ as $r$ increases to larger and larger values, and this profile induces the topological behavior which we will show below.  

%%%%%%%%%%%%%%%%%%%%%%%%%%%%%%%%%%%%%%%
%%%%%%%%%%%%%%%%%%%%%%%%%%%%%%%%%%%%%%%%
%%%%%%%%%%%%%%%%%%%%%%%%%%%%%%%%%%%%%%%
%%%%%%%%%%%%%%%%%%%%%%%%%%%%%%%%%%%%%%%

The corresponding energy density is given by
\be
\rho_{s}(r)=\frac{2}{(1-s)^2}r^{-2(3-s)/(1-s)}e^{-2r^{-2/(1-s)}} ,
\ee
and the total energy has the form
\be
E_s=\frac{\pi}{2(1-s)}.
\ee

We can calculate the mean matter radius introduced in \cite{7}. In this case we get
\be  
\bar{r}_s=2^{\frac12(1-s)}\,\Gamma\left(\frac{3+s}{2}\right).
\ee
It decreases as $s$ increases, and it is always smaller than it is for the standard vortex-like structure investigated in \cite{7}; in particular, for $s=0$ it is $\bar{r}=1.2533$, smaller than in the standard case of \cite{7}, which gives $\bar{r}=1.3561$. Moreover, one can use Eqs.~\eqref{q} and \eqref{T} to calculate the skyrmion number in this case; the result is $Q=1/2$. 

%%%%%%%%%%%%%%%%%%%%%%%%%%%%%%%%%%%%%%%%%
\subsection{Other topological solutions}

\begin{figure}[t!]
\centerline{\includegraphics[scale=0.4]{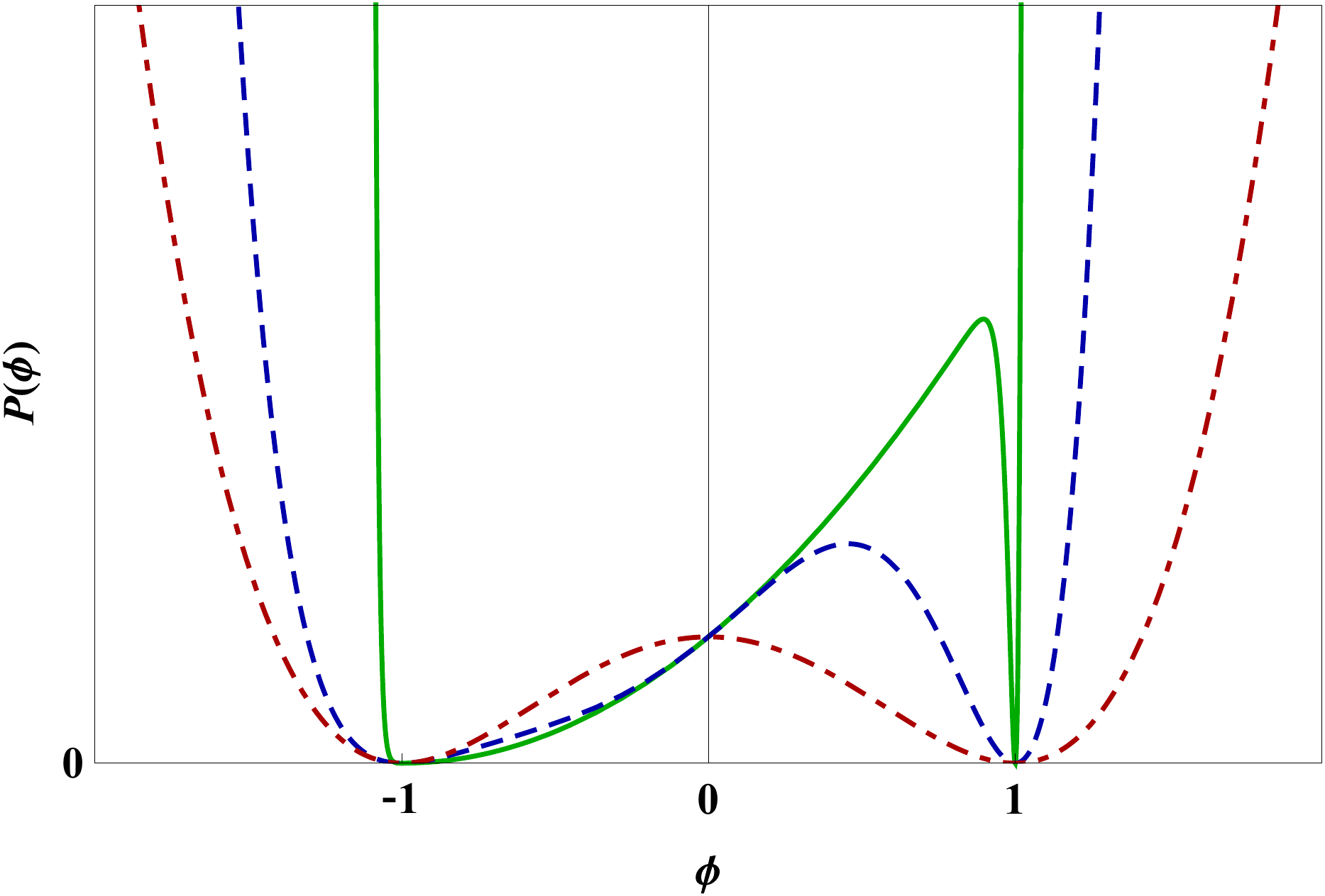}}
\caption{The polynomial \eqref{p3}, depicted for $s=0$, and for $n=1,3$ and for $n\to\infty$, with dot-dashed (red), dashed (blue) and solid (green) lines, respectively.}\label{fig5}
\end{figure}

The third model is described by
\be\label{p3}
P(\phi)=\frac1{(1-s)^2}(1+\phi)^2(1-\phi^n)^2
\ee
where $s\in[0,1)$, and $n=1,3,5,\cdots$. This polynomial has two minima at $\phi_\pm=\pm 1$, irrespective of the values of $s$ and $n$, and a local maximum in between $[0,1]$, whose value depends on $n$, and increases as $n$ increases to larger and larger values. We depict \eqref{p3} in Fig.~\ref{fig5} for $s=0$ and for three distinct values of $n$, to illustrate how the polynomial appears as $n$ increases to very larger values. We note that for $n=1$ we get to polynomial already studied in \cite{7}, so we do not consider it here. This model first appeared in \cite{as}, and there it was used to describe asymmetric structures capable of generating braneworld scenario having asymmetric behavior. 

%%%%%%%%%%%%%%%%%%%%%%%%%%%%%%%%%%%%%%%
%%%%%%%%%%%%%%%%%%%%%%%%%%%%%%%%%%%%%%%%
%%%%%%%%%%%%%%%%%%%%%%%%%%%%%%%%%%%%%%%
\begin{figure}[t!]
\centerline{\includegraphics[scale=0.4]{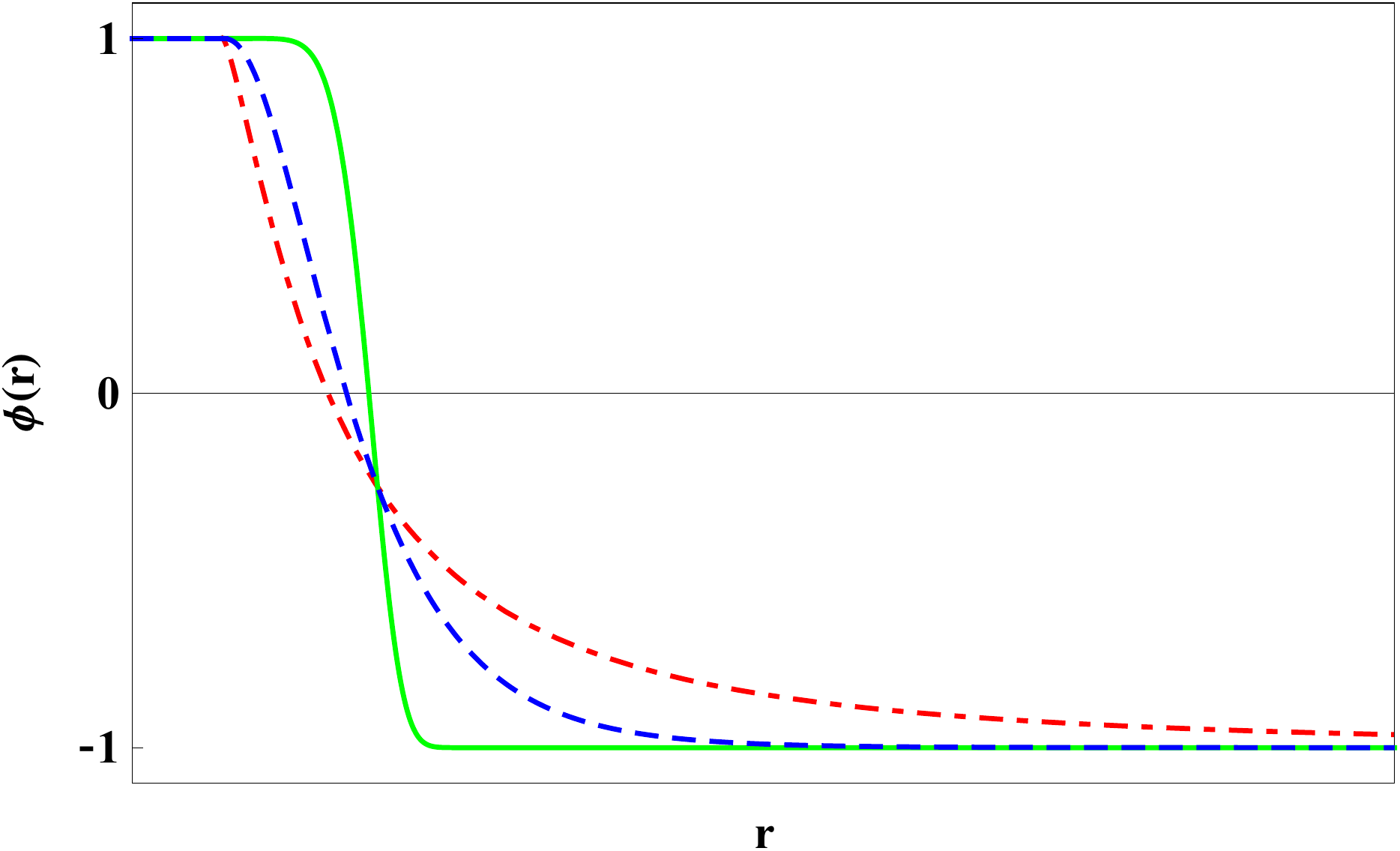}}
\caption{The solution \eqref{soln} for the model \eqref{p3} in the limit $n\to\infty$, depicted for $s=0.3, 0.6$ and $0.9$, with dot-dashed (red), dashed (blue) and solid (green) lines, respectively.}
\label{fig6}
\end{figure}
%%%%%%%%%%%%%%%%%%%%%%%%%%%%%%%%%%%%%%%

The equation of motion in this case becomes
\begin{eqnarray}
&&r^2\frac{d^2\phi}{dr^2}+ r \frac{d\phi}{dr} - \frac{(1 +\phi)(1-\phi^n)^2}{(1-s)^2} + \nonumber \\
 &&\qquad\qquad +\frac{n\phi^{n-1}(1+\phi)^2 (1-\phi^n)}{(1-s)^2}  = 0,
\end{eqnarray}
and here one searches for solutions that connect the two minima, $\phi_\pm=\pm1$. The above equation can be solved analytically in the limit $n\to\infty$ to give, 
\begin{eqnarray}\label{soln}
\phi_s(r)=\left\lbrace
\begin{array}{ll}
    \ \displaystyle 1, \quad \quad \quad \quad \quad \ \ \ 0 < r \leq1; \\ 
    \ \\
    \  \displaystyle \frac{2- r^{1/(1-s)}}{r^{1/(1-s)}}, \quad r > 1. \\ 
\end{array}
\right.
\end{eqnarray}
It is depicted in Fig.~\ref{fig6} for three distinct values of $s$. The energy density is
\begin{eqnarray}
\rho_s(r)=\left\lbrace
\begin{array}{ll}
    \ \displaystyle 0,  \quad \quad \quad \quad \quad \quad \ \ \ \  0 < r \leq 1; \\ 
    \ \\
    \  \displaystyle \frac{4}{(1-s)^2}\, r^{-2(2-s)/(1-s)}, \quad r > 1. \\ 
\end{array}
\right.
\end{eqnarray}
which leads to the total energy
\be
E_s=\frac{4\pi}{(1-s)}.
\ee

In this model, the mean matter radius is
\be  
{\bar{r}}_s=\frac{2}{1+s}.
\ee
It is always greater than it is in the standard case; in particular, for $s=0$ is has the highest value, ${\bar{r}}=2.0000$;
this is to be contrasted with the value ${\bar{r}}=1.1781$ that one gets in the standard case described in \cite{7}. In this case, if one uses Eqs.~\eqref{q} and \eqref{T} it is easy to calculate the skyrmion number, giving $Q=1$. 

%%%%%%%%%%%%%%%%%%%%%%%%%%%%%
\subsection{Stability}

We now study stability of the solutions found above. We consider stability against spherically symmetric deformations, taking $\phi=\phi(r)+\epsilon\,\eta(r)$, with $\epsilon$ being very small real and constant parameter. This allows that we write the total energy in the form
\be
E_\epsilon =E_0+\epsilon E_1+\epsilon^2 E_2+\cdots
\ee
where $E_n, n=1,2,...$ is the contribution to the energy at order $n$ in the small parameter $\epsilon$. We can use the equation of motion to show that $E_1 = 0$ and we obtain that 
\be \label{zeromode} 
 \eta(r)= A\, \text{exp}\left(\int dr\frac{1}{r^2}\frac{P'(\phi)}{\phi'(r)}\right) .
\ee
This is the zero mode, where $A$ is constant of normalization, $P'(\phi)=dP(\phi)/d\phi$ and $\phi'(r)=d\phi_s(r)/dr$.

We consider the model \eqref{p2}, and from (\ref{zeromode}) it is possible to write
\be  
\eta(r)=A_n \,\text{cos}^{n-1}\left(\text{ln}(r)\right)\sin{\left(\text{ln}(r)\right)}
\ee
in the interval $e^{-\pi/2} \leq r \leq e^{\pi/2}$; also, it vanishes outside this compact interval. Here,
$A_n$ is constant of normalization. We can prove that $E_2=0$, $E_3=0$ and $E_4$ has the form
\be  
E_4= -\frac{\pi^{3/2}(2+n) \Gamma(n-5/2)}{8n^2\Gamma(n-2)},
\ee
We see that $E_4$ is negative, and this proves that the spherically symmetric solution \eqref{phi2} is unstable against spherically symmetric fluctuations.

%%%%%%%%%%%%%%%%%%%%%%%%%%%%%%%%%%%%%%%%
\begin{figure}[t!]
\centerline{\includegraphics[scale=0.4]{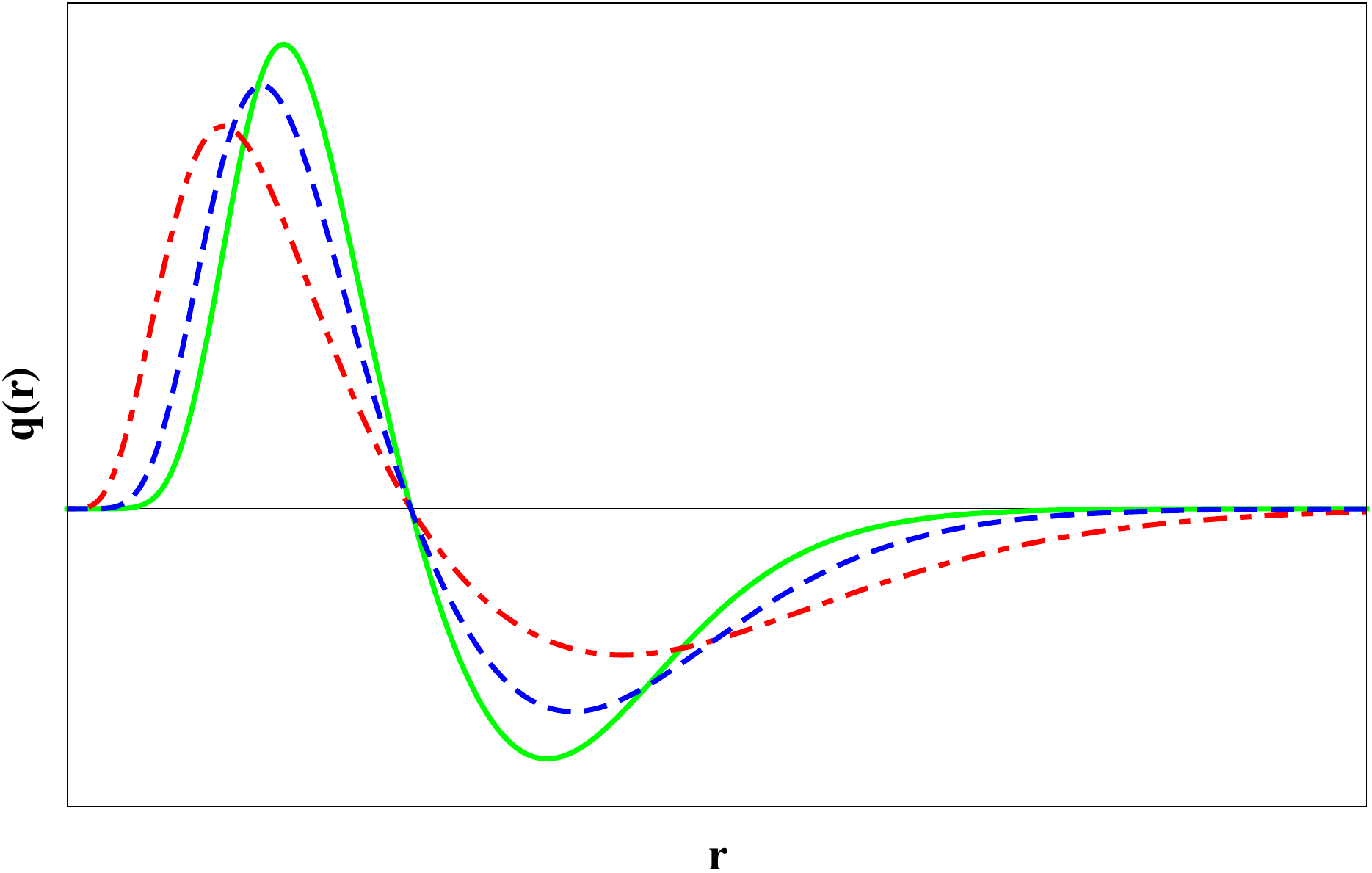}}
\caption{The non-topological charge density \eqref{topol1}-\eqref{topol2}, depicted for $n=3, 5$ and $7$, with dot-dashed (red), dashed (blue) and solid (green) lines, respectively.}
\label{fig7}
\end{figure}
%%%%%%%%%%%%%%%%%%%%%%%%%%%%%%%%%%%%%%%%

The next case is the model \eqref{p1}, and now the zero mode is given by
\be 
\eta(r)=A_s\, r^{-2/(1-s)}e^{-r^{-2/(1-s)}},
\ee
where $A_s$ is constant of normalization. We can prove that $E_2=0$, $E_3=0$ and $E_4$ has the form
\be 
E_4= \frac{3\pi}{16(1-s)}.
\ee
We see that it is positive, $E_4>0$ for any $s\in[0,1)$. This proves that the solution $\phi(r)$ which we obtained in \eqref{phi1} is stable against spherically symmetric fluctuations.

The last case is the model \eqref{p3}. It has analytical solution in the limit $n\to\infty$, given by Eq.~\eqref{soln}. We follow as we did before, to show that the solution is also stable against radial fluctuations. The zero mode is
\be  
\eta(r)=A\, r^{-1/(1-s)}, \quad r>1
\ee
where $A$ is normalization constant. The energy is such that $E_1=0$, and $E_2$ has the form
\be  
E_2=\frac{\pi}{1-s}
\ee
which is positive, indicating stability.

%%%%%%%%%%%%%%%%%%%%%%%%%%%%%%%%%%%%%%%%
\begin{figure}[t!]
\centerline{\includegraphics[scale=0.4]{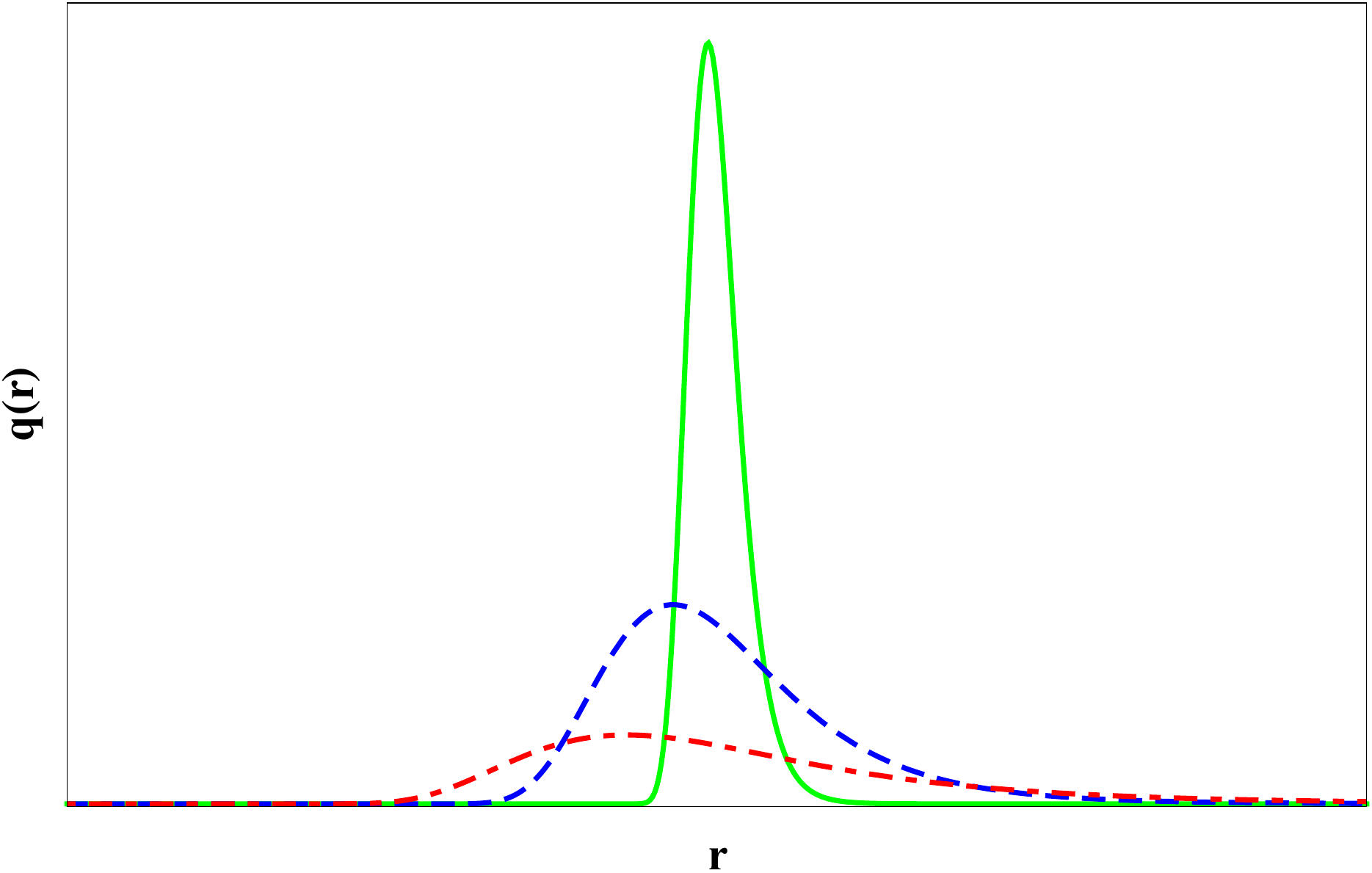}}
\caption{The topological charge density \eqref{topolo1} and \eqref{topolo2}, depicted for $s=0.3, 0.6$ and $0.9$, with dot-dashed (red), dashed (blue) and solid (green) lines, respectively.}
\label{fig8}
\end{figure}
%%%%%%%%%%%%%%%%%%%%%%%%%%%%%%%%%%%%%%%%

%%%%%%%%%%%%%%%%%%%%%%%%%%%%%%%%%%%%%%%%
\begin{figure}[t!]
\centerline{\includegraphics[scale=0.4]{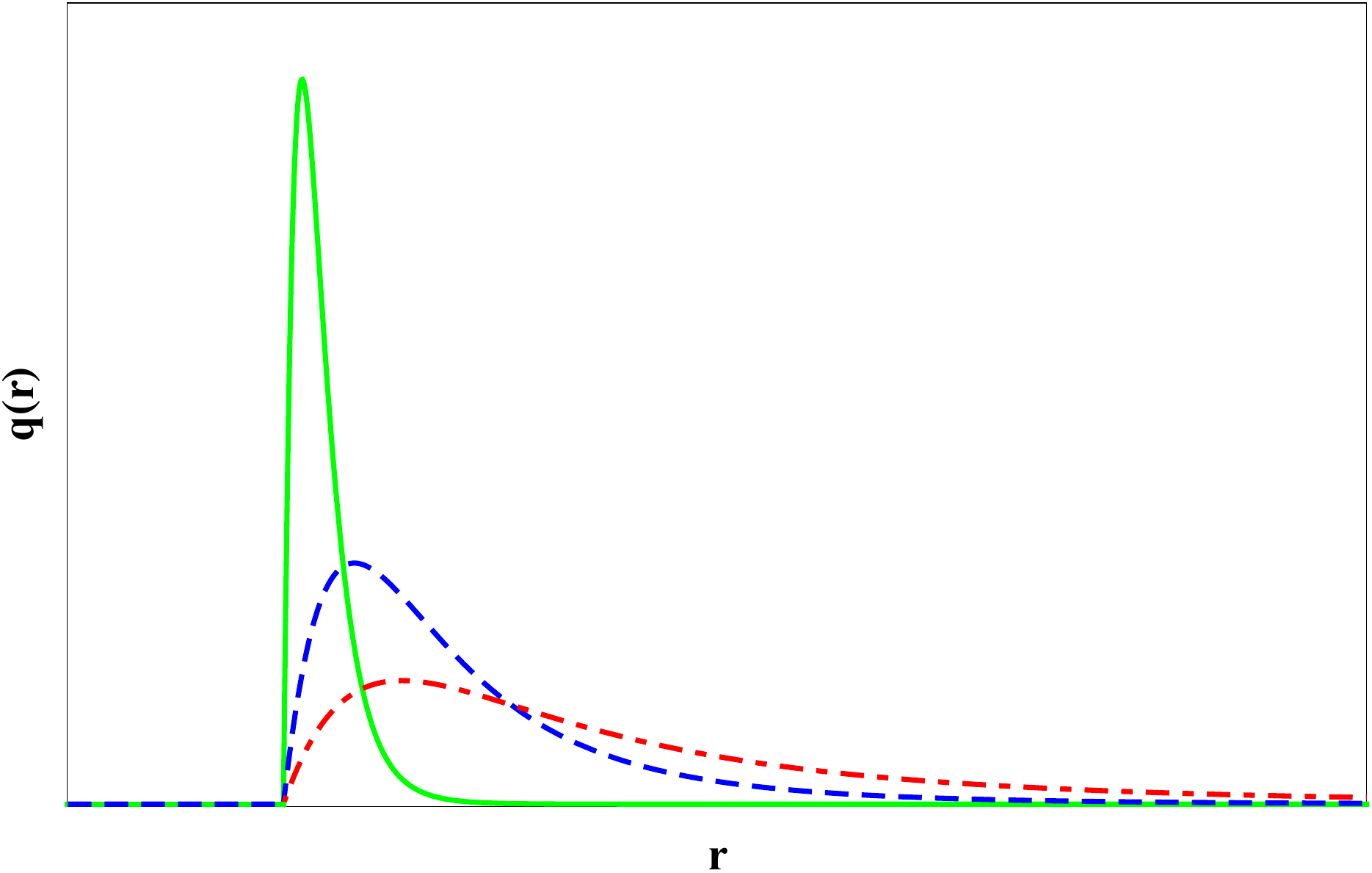}}
\caption{The topological charge density \eqref{qq3} and \eqref{qq3a}, depicted for $s=0.3, 0.6$ and $0.9$, with dot-dashed (red), dashed (blue) and solid (green) lines, respectively.}
\label{fig9}
\end{figure}
%%%%%%%%%%%%%%%%%%%%%%%%%%%%%%%%%%%%%%%%

%%%%%%%%%%%%%%%%%%%%%%%%%%%%%%%%%%
\section{Skyrmion number}

Let us now deal with the magnetization, as introduced in the beginning of the previous Section. It is possible to use \eqref{Q} to get the topological charge density in the form
\be\label{tcd}
q(r)= \frac{r}{2}{\bf M}\cdot\partial_x{\bf M}\times\partial_y{\bf M},
\ee
where {\bf M} is the magnetization. After taking \eqref{M}, \eqref{q}, and \eqref{T} one gets

\be\label{ttcd1}
q(r)=-\frac{\pi}{4}\cos\Theta(r)\frac{\partial\phi(r)}{\partial r}.
\ee

We now apply this procedure to the models investigated above. 
For the model $(\ref{p2})$ we have that
\be\label{topol1}
q(r)=q_0(r)\sin{\left(\frac{\pi}{2}\text{cos}^n\left(\text{ln}(r)\right)\right)},
\ee
where we used $\delta=\pi/2$ and $q_0(r)$ is given by
\be\label{topol2}
q_0(r)=\frac{\pi n\text{cos}^{n-1}\left(\text{ln}(r)\right)\sin{\left(\text{ln}(r)\right)}}{4r}
\ee
which we depict in Fig. \ref{fig7} for three distinct values of $n$.

%%%%%%%%%%%%%%%%%%%%%%%%%%%%%%%%%%%%%%%%
\begin{figure}[t!]
\centerline{\includegraphics[scale=0.125]{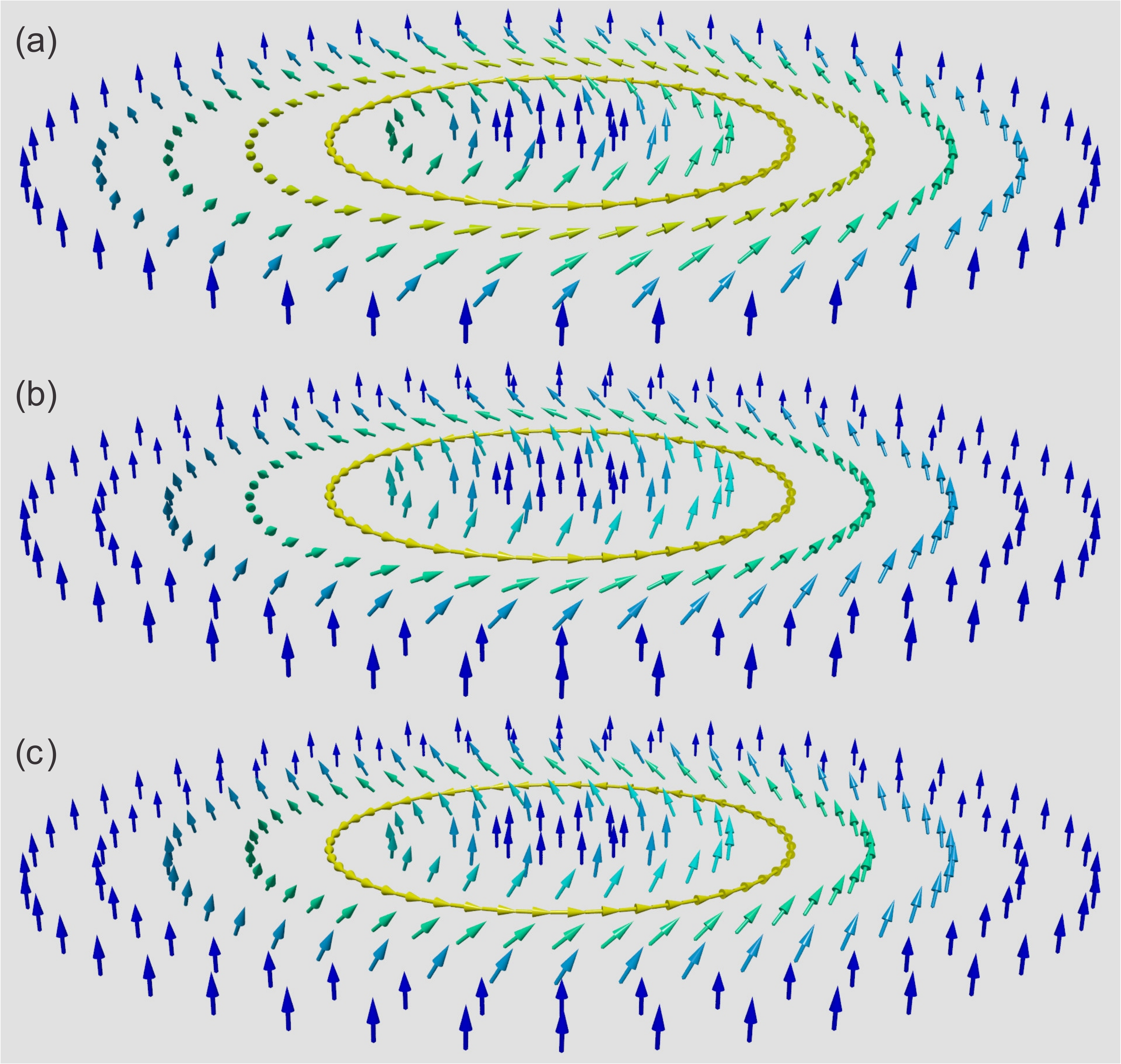}}
\caption{(Color online) The non-topological structure with skyrmion number $Q=0$ which is controlled by the model (\ref{p2}), depicted for $n=3, 5$ and $7$ in the (a), (b) and (c) panels, respectively.}
\label{fig10}
\end{figure}
%%%%%%%%%%%%%%%%%%%%%%%%%%%%%%%%%%%%%%%%

%%%%%%%%%%%%%%%%%%%%%%%%%%%%%%%%%%%%%%%%
\begin{figure}[t!]
\centerline{\includegraphics[scale=0.22]{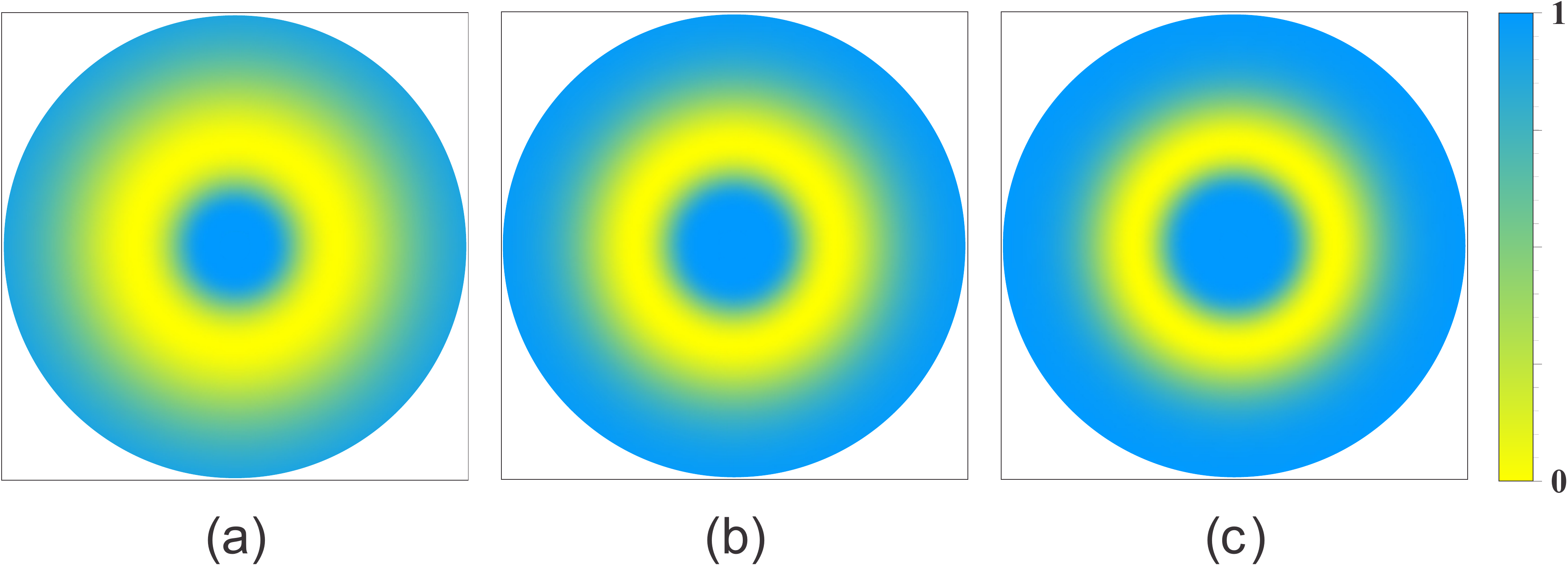}}
\caption{(Color online) Alternative way to depict the non-topological structure controlled by the model (\ref{p2}), for $n=3,5$ and $7$, in the (a), (b) and (c) panels, respectively.}
\label{fig11}
\end{figure}
%%%%%%%%%%%%%%%%%%%%%%%%%%%%%%%%%%%%%%%%

For the model (\ref{p1}) we use
$\delta = \pi/2$ and get 
\be\label{topolo1}
q(r)=q_0(r)\sin{\left(\frac{\pi}{2}e^{-r^{-2/(1-s)}}\right)},
\ee
where 
\be\label{topolo2}
q_0(r)=\frac{\pi r^{-(3-s)/(1-s)} e^{-r^{-2/(1-s)}}}{2(1-s)}
\ee
which we depict in Fig. \ref{fig8} for three distinct values of $s$. 

For the model \eqref{p3} the topological charge density becomes, using $\delta=0$,
\be\label{qq3}
q(r)=q_0(r)\cos{\frac{\pi}{2} \left( \frac{2-r^{1/(1-s)}}{r^{1/(1-s)}}\right) },
\ee
where
\be\label{qq3a}
q_0(r)=\frac{\pi r^{-(2-s)/(1-s)}}{2(1-s)},
\ee
which is depicted in Fig.~\ref{fig9} for three distinct values of $s$.

To get the skyrmion number or the topological charge we recall that 
\be\label{Q1}
Q = \int^\infty_0 drq(r),
\ee
where $q(r)$ is given by eq. (\ref{ttcd1}). We use this and the eqs.~\eqref{topol1} and (\ref{topol2}) to see that $Q=0$ in this case. The skyrmion number for the model (\ref{p2}) vanishes, so the model supports a non-topological compact structure. The Fig.~\ref{fig10} describes an array of vectors in the plane, which represents the helical magnetic excitation with vanishing skyrmion number. Alternatively, we can depict the magnetic structures using a continuum of color, taking blue for the magnetization pointing upward in the $\hat{z}$ direction, yellow for the magnetization vanishing along the $\hat{z}$ direction, and red for the magnetization pointing downward in the $\hat{z}$ direction. We use this to depict in Fig.~\ref{fig11} the non-topological solution with the same values of $s$ considered in Fig.~\ref{fig10}.

%%%%%%%%%%%%%%%%%%%%%%%%%%%%%%%%%%%%%%%%
\begin{figure}[t!]
\vspace{1cm}
\centerline{\includegraphics[scale=0.125]{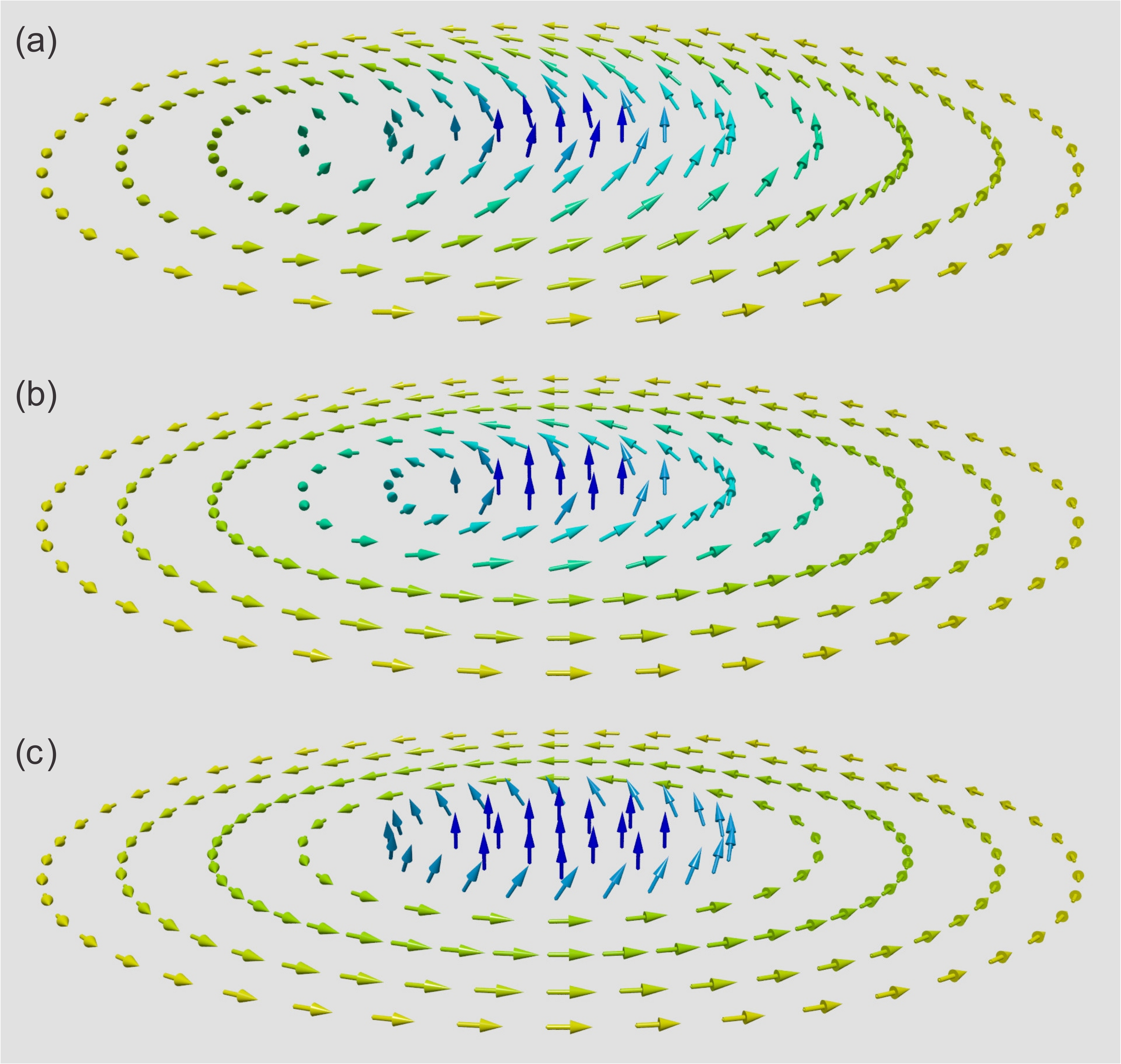}}
\caption{(Color online) The topological structure with skyrmion number $Q=1/2$ which is controlled by the model (\ref{p1}), depicted for $s = 0.3, 0.6$ and $0.9$ in the (a), (b) and (c) panels, respectively.}
\label{fig12}
\end{figure}
%%%%%%%%%%%%%%%%%%%%%%%%%%%%%%%%%%%%%%%%
%%%%%%%%%%%%%%%%%%%%%%%%%%%%%%%%%%%%%%%%
\begin{figure}[t!]
\centerline{\includegraphics[scale=0.22]{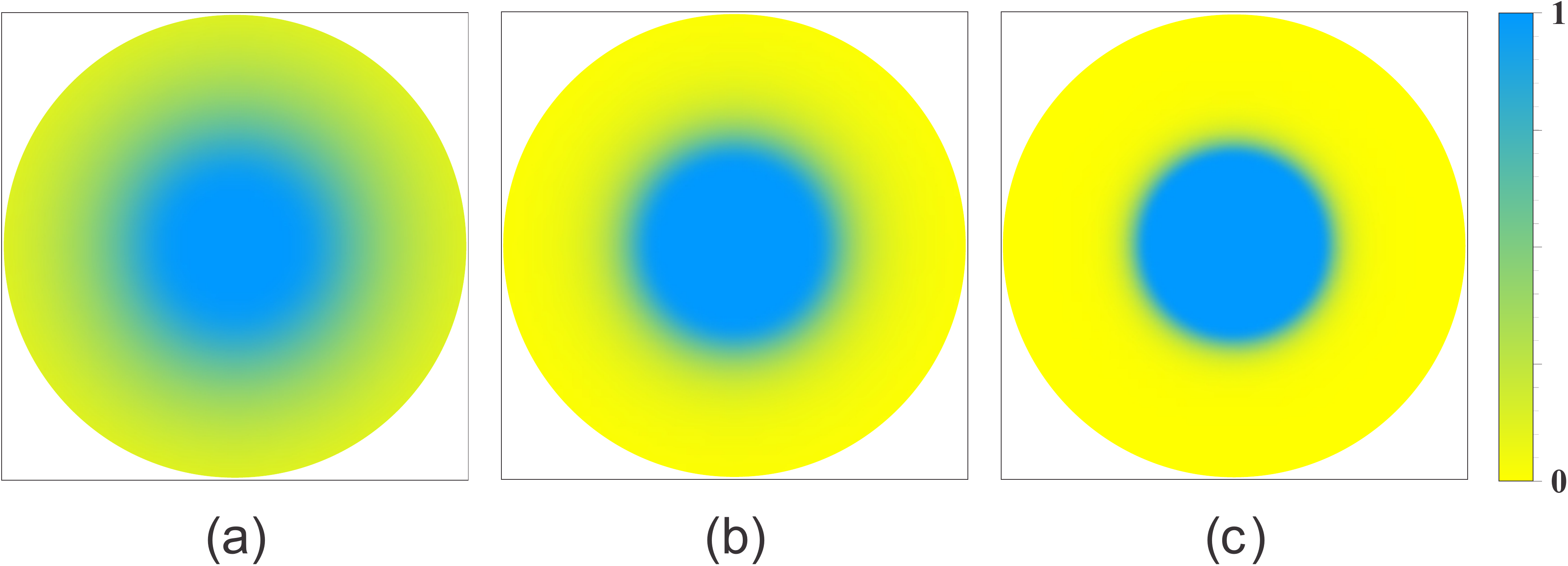}}
\caption{(Color online) Alternative way to depict the topological structure controlled by the model (\ref{p1}),
for $s = 0.3$, $0.6$ and $0.9$, in the (a), (b) and for (c) panels, respectively.}
\label{fig13}
\end{figure}
%%%%%%%%%%%%%%%%%%%%%%%%%%%%%%%%%%%%%%%%

We now consider \eqref{ttcd1} and use \eqref{topolo1} and \eqref{topolo2} to get that $Q=1/2$. In this case we are dealing with a skyrmion with skyrmion number $1/2$ which represents a topological structure of the vortex type. We depict some of these vortex structures in Fig.~\ref{fig12}. Also, in Fig.~\ref{fig13} we depict the same figures, but now using a continuum of colors, as we did before for the non-topological structure. We note that the structures behave standardly asymptotically, however, the field configuration vanishes rapidly, as $r$ diminishes toward the origin, so it represents a semi-compact vortex-like structure. 

%%%%%%%%%%%%%%%%%%%%%%%%%%%%%%%%%%%%%%%%
\begin{figure}[t!]
\vspace{1cm}
\centerline{\includegraphics[scale=0.126]{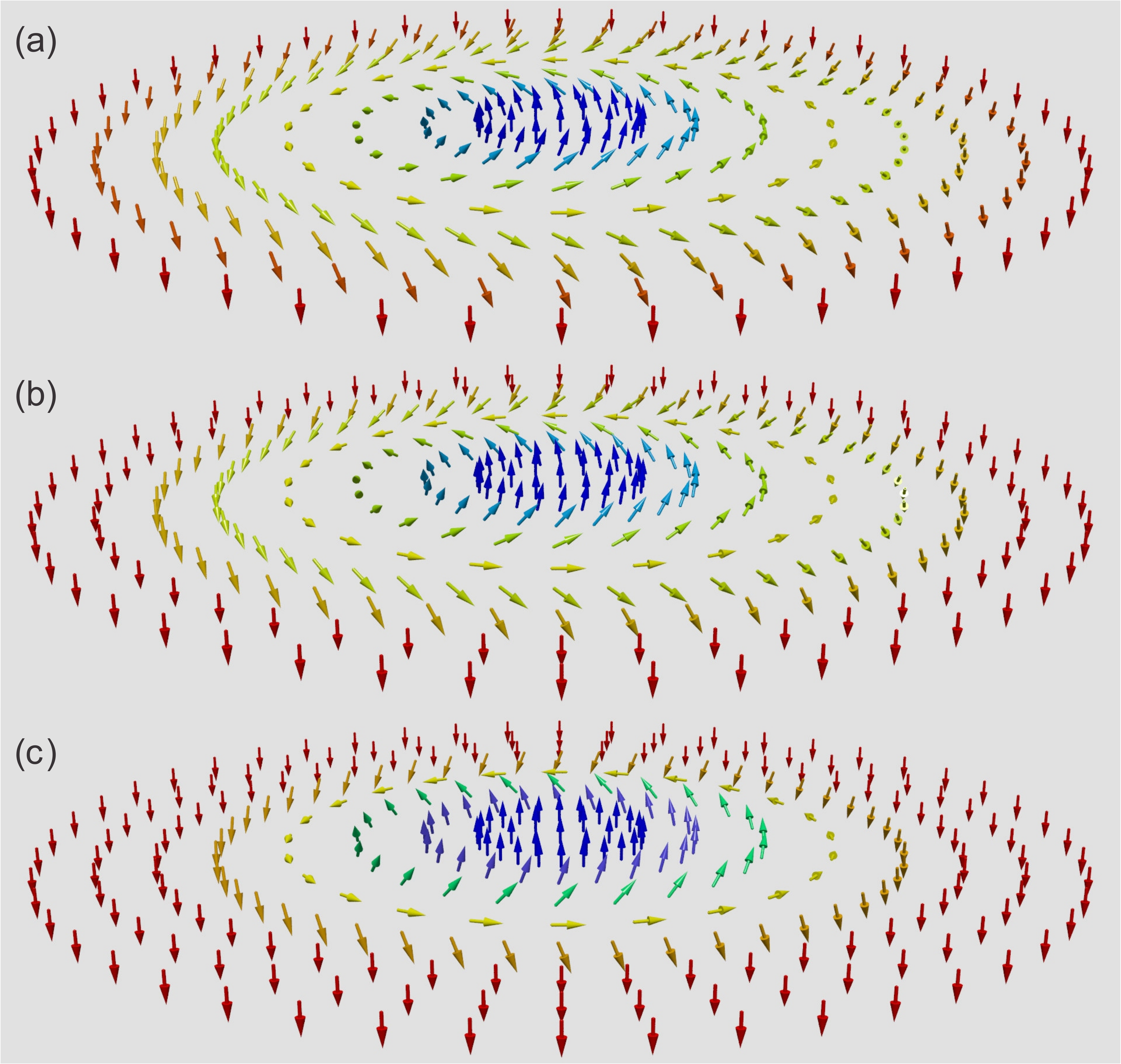}}
\caption{(Color online) The topological structure with skyrmion number $Q=1$ which is controlled by the model (\ref{p3}), depicted for $s = 0.3, 0.6$ and $0.9$ in the (a), (b) and (c) panels, respectively.}
\label{fig14}
\end{figure}
%%%%%%%%%%%%%%%%%%%%%%%%%%%%%%%%%%%%%%%%
%%%%%%%%%%%%%%%%%%%%%%%%%%%%%%%%%%%%%%%%
\begin{figure}[t!]
\centerline{\includegraphics[scale=0.22]{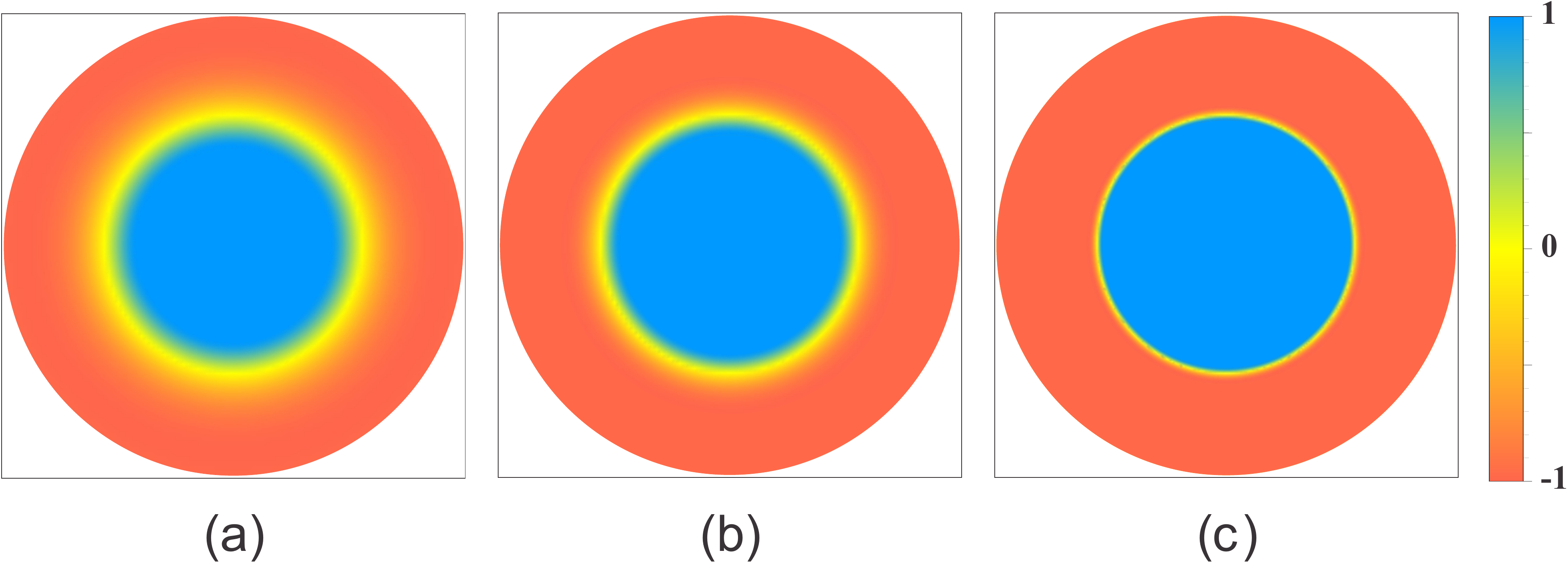}}
\caption{(Color online) Alternative way to depict the topological structure controlled by the model (\ref{p3}),
for $s = 0.3$, $0.6$ and $0.9$, in the (a), (b) and for (c) panels, respectively.}
\label{fig15}
\end{figure}
%%%%%%%%%%%%%%%%%%%%%%%%%%%%%%%%%%%%%%%%

The last model is controlled by \eqref{p3}, so we use \eqref{qq3} and \eqref{qq3a} to get that $Q=1$. The solutions then represent topological structures of the skyrmion type with unit skyrmion number. We depict some of them in Figs.~\ref{fig14} and \ref{fig15}, using an array of vectors and a continuum of colors, respectively. We note that for $Q=1$, the skyrmion-like structures behave as standard structures as $r$ increases to larger and larger values; however, for $r$ small, the spin structures quickly become uniform, unveiling its semi-compact character around its inner core, in a way similar to the case of the vortex-like configurations.  

%%%%%%%%%%%%%%%%%%%%%%%%%%%%%%%%%%
\section{ENDING COMMENTS} 

We described three types of planar, localized and spherically symmetric structures, one of them having non-topological behavior, and the other being of topological nature. The non-topological structure  can be used to describe compact structure, having energy density localized in a compact region in the plane. It is, however, unstable against spherically symmetric fluctuations and should be further studied to be stabilized. 

On the other hand, the topological structures are stable under spherically symmetric fluctuations, and have skyrmion number $1/2$ and $1$, so they are capable of modeling vortex- and skyrmion-like textures, respectively. 
They have asymptotic behavior similar to the standard vortex and skyrmion configurations. However, since the associated fields quickly vanish as one approaches the center of the configurations, they indicate a compact behavior at the corresponding cores. Thus, they behave as semi-compact structures. Interestingly, they engender a single parameter, which can be used the model the internal configurations and describe how the associated spin textures behave around their central region. The results seem to be of current interest since it provides a way to control the magnetic profile inside the localized structures. In this sense, the two topological models unveil a route that helps us tayloring the core of the vortex- and skyrmion-like textures at the nanometric scale.

This work is partially supported by CNPq, Brazil. DB acknowledges support from grants 455931/2014-3 and 306614/2014-6, and EIBR acknowledges support from grant 160019/2013-3.

%%%%%%%%%%%%%%%%%%%%%%%%%%%%%%%%%

\end{document}